\documentclass[aps,prd,eqsecnum,nofootinbib,preprintnumbers,showpacs]{revtex4}
\def\sqr#1#2{{\vcenter{\vbox{\hrule height.#2pt\hbox{\vrule
width.#2pt height#1pt \kern#1pt\vrule width.#2pt}\hrule height.#2pt}}}}
 
\usepackage{latexsym}
\usepackage{bm}
\begin{document}

\thispagestyle{empty}

\rightline{\large\baselineskip20pt\rm\vbox to20pt{
\baselineskip14pt
\hbox{YITP-05-03}}}
\vspace{1mm}

\vskip8mm
\begin{center}
{\large  \bf
Kaluza-Klein gravitons are negative energy dust in brane cosmology}
\end{center}

\def\thefootnote{\fnsymbol{footnote}}
\def\beq{\begin{equation}}
\def\eeq{\end{equation}}

\begin{center}
{\large
Masato Minamitsuji$^{1,2}$\,,\,
Misao Sasaki$^2$ }
\end{center}
\begin{center}
{\em
$^1$Department of Earth and Space Science, Graduate School of Science,
\\
Osaka University, Toyonaka 560-0043, Japan}
\end{center}
\begin{center}
{\em
$^2$Yukawa Institute for Theoretical Physics,
\\
 Kyoto University, Kyoto 606-8502, Japan}
\end{center}

\begin{center}
 {\large David Langlois$^{3,4}$}
\end{center}
\begin{center}
{\em 
$^3$ APC
\footnote{UMR 7164 (CNRS, Universit\'e Paris 7, CEA, 
Observatoire de Paris)} (Astroparticules et Cosmologie),\\
 11 Place Marcelin Berthelot, F-75005 Paris, France}
\end{center}
\begin{center}
{\em
$^4$ GRECO, Institut d'Astrophysique de Paris,
\\ 98bis Boulevard Arago,
 75014 Paris, France
}  
\end{center}


\begin{abstract}
We discuss the effect of Kaluza-Klein (KK) modes of bulk metric
perturbations on the second Randall-Sundrum (RS II) type brane
cosmology, taking the possible backreaction in the bulk and on the
brane into account.
 KK gravitons may be produced via quantum fluctuations during
a de Sitter (dS) inflating phase of our brane universe.
In an effective 4-dimensional theory in which one integrates out
the extra-dimensional dependence in the action, KK gravitons
are equivalent to massive gravitons on the brane
with masses $m>3H/2$, where $H$ represents the expansion rate of a dS brane.
Thus production of even a tiny amount of KK gravitons may eventually
have a significant impact on the late-time brane cosmology.
As a first step to quantify the effect of KK gravitons on the brane, 
we calculate the effective energy density and pressure for a single
KK mode. Surprisingly, we find that a KK mode behaves as cosmic dust
with a %
{\sl negative} energy density on the brane. We 
note that
the bulk energy density of a KK mode is positive definite
and there occurs no singular phenomenon in the bulk.
\end{abstract} 

\pacs{04.50.+h; 98.80.Cq}

\date{\today}

\maketitle

\makeatletter
\makeatother

\section{Introduction}
The idea that our Universe might be a brane embedded in a higher dimensional
bulk spacetime has attracted tremendous attention in the last few years. 
A particularly attractive framework, especially from a gravitational 
point of view, is the so-called second Randall-Sundrum (RS II)
scenario, where our brane-universe is embedded in an Anti-de Sitter
(AdS) five-dimensional bulk spacetime \cite{Randall:1999vf}. 
If the brane is endowed with a (positive) tension, tuned  with respect
to the (negative) bulk cosmological constant $\Lambda_{5}\equiv
-6/\ell^2$, then, as shown by Randall and Sundrum, the geometry on the
brane is Minkowski and the gravity felt on the brane is similar to
standard 4D gravity on scales $r\gg \ell$ \cite{Garriga:1999yh, Kanno:2002ia}.

A considerable amount of work has also been devoted to the cosmological 
extension of the RSII model, in order to describe the 
cosmological behavior of a brane in this framework (see
\cite{Langlois:2002bb, Brax:2002yh, Maartens:2003tw, Brax:2004xh}
for reviews).
 A significant result in this perspective has been the realization
that the Friedmann equations
must be modified when the cosmological energy density becomes of the 
order of the brane tension. Another difference with the standard Friedmann 
equation is the presence of an additional term, usually called dark radiation, 
which represents the influence of the bulk geometry on the brane cosmological
evolution. The dark radiation term can be related, from the bulk 
point of view, to a five-dimensional gravitational mass.

All the fundamental results just discussed have been obtained by 
assuming from the start an exact cosmological symmetry, by which we mean 
that the bulk spacetime is supposed to be foliated by homogeneous and isotropic
three-surfaces. At all instants in its history the brane is supposed 
to coincide with one of these symmetric three-surfaces. However, things are 
more complicated if one considers a more realistic framework where
 perturbative deviations from homogeneity
and isotropy are allowed.
Such perturbations must be taken into account, for example,
if one wishes to confront the 
predictions of brane cosmology with the high-precision measurements of 
the CMB anisotropies. 

Even if one is interested only in homogeneous and isotropic cosmology,
the existence of perturbations can affect the homogeneous
evolution. An interesting example is the question of dark
radiation. The presence of matter fluctuations on the brane generates
perturbations of the spacetime metric in the bulk. This process may be
regarded as the emission of bulk gravitons from the brane.
These bulk gravitons will on average contribute to the gravitational mass
in the bulk, and hence modify the evolution of the dark radiation on the
brane. That is, it no longer behaves like a free, conserved radiation as
in the strictly symmetric case.
Detailed analytical and numerical calculations of the
effect of the bulk graviton generation were performed for a 
radiation-dominated brane universe
\cite{hm,Langlois:2002ke,Langlois:2003zb,Leeper:2003dd,Minamitsuji:2003at,
Moss:2004dp, Chang:2004xs}.

In the present work, we consider a similar, but slightly different
problem. We study the impact of the bulk metric perturbations
which are generated in the bulk (or present from the beginning)
 on the homogeneous evolution of the brane. 
 The bulk metric perturbations are naturally produced via
quantum fluctuations during brane
 inflation~\cite{Langlois:2000ns}, 
and we will concentrate in this work mainly on a de Sitter (dS)
brane. From the 4-dimensional point of view,
 the bulk metric perturbations %
  can be decomposed into %
  a massless 
zero mode and an infinite number of Kaluza-Klein (KK) modes with
effective mass  $m>3H/2$, where $H$ is the expansion rate of the dS brane.
The time evolution of the zero mode is similar to the standard four
dimensional perturbations although its amplitude, as determined by 
the vacuum quantum fluctuations, depends on the energy
scale of the dS brane expansion. 
In contrast, the squared-amplitude of KK modes on the brane decays as
$a^{-3}$ and thus becomes rapidly negligible during brane inflation.
However, after brane inflation the background energy density in a
radiation-dominated Friedmann-Lema\^\i tre-Robertson-Walker (FLRW)
era decays as $a^{-4}$, hence the massive modes of
gravitons may affect the late-time cosmological evolution of the brane.
Here it should be noted that the concept of KK modes, which assumes
the separation of variables with respect to the fifth dimensional
coordinate, is only approximately defined in general. 
It is an important but longstanding problem to quantify 
the effect of these approximately defined KK modes on the brane
evolution.

In this paper, in order to discuss the cosmological impact of KK gravitons
on the brane cosmology, we derive the effective stress-energy of a KK
mode on a separable (e.g., a dS brane) background, taking possible
backreaction in the bulk and on the brane into account.
Then we extrapolate %
our result to a FLRW cosmological background on which a KK
mode can be approximately defined. 
We show that a sufficiently massive KK mode, which may constitute
a non-negligible, if not dominant, fraction of the contribution 
of all the KK mode, when they are summed up, behaves as cosmic dust, 
which is consistent with the linear perturbations, but
the effective energy density %
is negative.

This paper is organized as follows.
In Section~II, we discuss the case of a massless, minimally coupled
scalar field because the situation is similar to the tensor case but
simpler. We find that a massive KK mode behaves like %
 dust with negative energy density.
In Section~III, we turn to the main topic of this paper, namely,
the backreaction of the KK gravitons on %
the cosmology of the brane.
We find again that a KK graviton mode behaves as negative
energy dust. 
In Section~IV, we consider our results from the bulk point of view,
and discuss its impact on the cosmological evolution.
In Section~V, we summarize our results.
Some useful formulae are given in two appendices. 
In Appendix~A, the components of the bulk curvature tensor up to 
second order in the  metric perturbations are given.
In Appendix~B, the computational rules for averaging tensor components 
that are quadratic in the metric perturbation are given.

\section{The case of a bulk scalar field}

Before tackling the main subject of this paper, 
that is the backreaction of KK gravitons, it is instructive to 
 discuss the case of a massless, minimally coupled scalar field,
because the behavior of its perturbations is quite similar to the 
KK gravitons but it is much simpler to analyze \cite{Himemoto:2000nd}.

We thus consider a homogeneous scalar field and assume  
its amplitude $\phi$ to be small so that its effect can  be treated
perturbatively: in particular, the backreaction of the 
scalar field on the metric will be of order  ${\cal O}(\phi^2)$. 

Eventually, we would like to discuss the backreaction for a 
general cosmological background. However, for the general case,
it turns out that the field equation (either for the scalar field
or, later, for the gravitons) is not separable and the notion of a
 KK mode cannot be well defined. 
The separability property is satisfied only for two limiting cases.
One is the case of a de Sitter brane.
In this case, the brane is exponentially expanding with a constant Hubble
rate $H$ and one finds a mass gap $\Delta m=3H/2$
between the zero mode and KK modes. Thus
 the continuum of KK modes starts above the mass $3H/2$.
The other case is a low energy cosmological brane, in which case the 
dependence on the extra dimension can be approximated by the profile 
obtained for a static brane, i.e. the RS brane.

\subsection{Einstein scalar theory in the bulk}

We start from the five-dimensional action which consists of
 the Einstein-Hilbert term, a cosmological constant $\Lambda_{5}$
 and a bulk scalar field, complemented by the four-dimensional action for the brane:
\begin{eqnarray} 
       S=  \frac{1}{2\kappa_{5}^2}
           \int d^5 x \sqrt{-g} \Bigl(
               {}^{(5)}R - 2  \Lambda_{5}
                        \Bigr)
           +\int d^5 x \sqrt{-g} 
                        \Bigl(
                       -\frac{1}{2}g^{ab}
                         \partial_{a} \phi
                         \partial_{b} \phi
                        -V(\phi)
                        \Bigr)  
           +\int d^4 x \sqrt{-q} \Bigl(
                         -\sigma 
                         + {\cal L}_{m}
                          \Bigr)\,,
                          \label{action}
\end{eqnarray}
where $q$ is the determinant of the induced metric on the brane,
which we denote by $q_{\alpha\beta}$, and 
${\cal L}_{m}$ is the Lagrangian density of the matter confined on the brane.
The Latin indices $\{a,b,\cdots\}$ and the Greek indices
$\{\alpha,\beta,\cdots\}$ are used for tensors defined in the bulk and on
the brane, respectively. We will assume that the brane 
tension on the brane is tuned %
to its RS value so that  $\kappa_{5}^4 \sigma^2=-6\Lambda_{5} $.
We also take a  constant bulk potential
\begin{eqnarray}
V(\phi)= V_0>0\,,
\end{eqnarray}
so that the scalar field is effectively massless. 

We consider
 backgrounds given by a fixed value of the scalar
field which we choose $\phi=0$. For a non-zero $V_0$, one has a
de Sitter brane background, which will be discussed in subsection~B below.
For $V_0=0$, one has a low energy cosmological brane, discussed in 
subsection~C.

The field equation for the bulk scalar field is linear and given by 
\begin{eqnarray}
\Box_{5}\, \phi =0.
\end{eqnarray}
Since we consider a background configuration 
with $\phi=0$, the solution of the above 
equation can be seen as a perturbation. This perturbation will 
induce a bulk energy-momentum tensor, of order ${\cal O}(\phi^2)$, 
which embodies the backreaction of the scalar field on the metric.
This is the effect we wish to calculate explicitly. 

The variation of the action (\ref{action}) yields the five-dimensional Einstein
equations
\begin{eqnarray}
    {}^{(5)}G_{ab}  +\Lambda_{5}g_{ab} 
         = -\kappa_{5}^2 V_0 g_{ab}
          +  \kappa_{5}^2 {\cal T}_{ab} 
          + \bigl(
             -\sigma q_{ab}
             +\tau_{ab}
            \bigr)
            \delta(y-y_0) 
\label{scalarbulk}
\end{eqnarray}
where we have implicitly assumed a coordinate system in which the
brane stays at a fixed location $y=y_0$ and where 
\begin{eqnarray}
\tau_{\alpha\beta}
        = \frac{2}{\sqrt{-q}}
          \frac{\delta}
               {\delta q^{\alpha\beta}}
               \Bigl(
                \sqrt{-q} {\cal L}_{m}
               \Bigr)\, 
\end{eqnarray}
represents the energy-momentum tensor of matter confined on the brane.
The  stress energy tensor of the bulk scalar
field, {\it not} including the constant potential $V_0$, is given by 
\begin{eqnarray}
   {\cal T}_{ab}=\phi_{,a}\phi_{,b}
          -\frac{1}{2}g_{ab}
              g^{cd}\phi_{,c}\phi_{,d}\,.
\end{eqnarray}
It is useful to consider the projection of the gravitational equations 
on the brane\cite{Shiromizu:1999wj}. Taking into account the bulk
energy-momentum tensor, one finds 
\begin{eqnarray}
   {}^{(4)}G^{\alpha}{}_{\beta}
  =-\frac{1}{2}\kappa_{5}^2 V_0 \delta^{\alpha}{}_{\beta}
  +\frac{1}{6} \kappa_{5}^4\sigma \tau^{\alpha}{}_{\beta}
  + \kappa_{5}^2 T^{(b)\alpha}{}_{\beta}
  -E^{\alpha}{}_{\beta}\,,
\end{eqnarray}
where
\beq
\label{Tb}
T^{(b)\alpha}{}_{\beta}
=  \frac{2}{3}
    \Bigl[
        \phi^{,\alpha}\phi_{,\beta}
        +\delta^{\alpha}{}_{\beta}
        \Bigl(
          \frac{3}{8}\phi_{,y}^2
         -\frac{5}{8}q^{\rho\sigma}\phi_{,\rho}
                                   \phi_{,\sigma}
        \Bigr)     \Bigr]\,,
\eeq
and $E_{\alpha\beta}$ is the projection on the brane of the bulk Weyl
tensor and is traceless by construction.  
If, in addition, one assumes the brane geometry to be homogeneous and isotropic
then the components of $E_{\alpha\beta}$ ( in an appropriate
coordinate system) reduce to $E^{t}{}_{t}$ and 
\beq
 E^{i}{}_{j}=-\frac{1}{3}\delta^{i}{}_{j}E^{t}{}_{t}.
\eeq
By using the four-dimensional Bianchi identities, and assuming that the brane 
matter content is conserved, one is able to express
the component $E^{t}{}_{t}$ in terms of the values on the brane of the 
bulk scalar field and its derivatives \cite{Himemoto:2000nd}:
\beq
 E^{t}{}_{t}   = 
      \frac{\kappa_{5}^2}{ a^4}
        \int^{t}_{t_0} dt' a^4 
         \Bigl( 
             \partial_{t} T^{(b)}{}^{t}{}_{t}
            +3\frac{\dot{a}}{a} T^{(b)}{}^{t}{}_{t}
            -\frac{\dot{a}}{a} T^{(b)}{}^{i}{}_{i}
         \Bigr)\,,
   \label{Ett}
\eeq

\subsection{KK mode on a de Sitter brane}

First, we consider the case of a de Sitter brane.
The bulk metric around a de Sitter brane can be expressed as 
\begin{eqnarray}
ds^2 =dy^2 + b^2(y)\gamma_{\mu\nu}dx^{\mu}dx^{\nu}\,,
\end{eqnarray}
where the warp factor $b(y)$ is given by
\begin{eqnarray}
  b(y)= H\ell \sinh({y}/{\ell})\,,
 \label{warp_dS}
\end{eqnarray}
and $\gamma_{\mu\nu}$ is the 4-dimensional de Sitter metric,
which may be expressed by using a flat slicing for simplicity:
\begin{eqnarray}
&& \gamma_{\mu\nu} dx^{\mu}dx^{\nu}
= -dt^2 +a^2(t)\delta_{ij}dx^i dx^j\,,
\nonumber \\
   && 
 a(t) =e^{Ht}\,,\quad 
 H^2=\frac{1}{6}\kappa_{5}^2 V_{0}\,. 
\label{dS_four}   
\end{eqnarray}
The brane is located at $y=y_0$ such that $b(y_0)=1$,
that is,
\begin{eqnarray*}
\sinh(y_0/\ell)=\frac{1}{H\ell}\,.
\end{eqnarray*}

In this geometry, the equation of motion for the scalar field  is
\begin{eqnarray}
    \frac{1}{b^4}
     \partial_y\Bigl(b^{4}\partial_y\phi\Bigr)
     -\frac{1}{b^2} 
             \Bigl(
               \ddot{\phi}
      +3H \dot{\phi}-\frac{1}{a^2}\mathop{\Delta}\limits^{(3)}\phi
             \Bigr)       
      =0\,.
\end{eqnarray}
This equations is separable and one can solve it by looking 
for a solution of the form $\phi=f(y)\varphi(t,x^i)$, with 
\begin{eqnarray}
  &&   \frac{1}{b^2}
     \partial_y\Bigl(b^{4}\partial_y f\Bigr)+m^2 f=0 \,, 
\nonumber \\
  && \ddot{\varphi}
       +3H \dot{\varphi}
       -\frac{1}{a^2}\mathop{\Delta}\limits^{(3)}\varphi
        +m^2\varphi
            =0\,.\label{dS_scalar}
\end{eqnarray}
The separation constant $m^2$ corresponds to the square of the 
 KK mass, as measured by an observer  on the brane.

Since there is no coupling between the brane and the bulk scalar field, 
the  boundary condition for  the scalar field at  the
brane location is simply $\partial_y\phi=0$, and therefore $\partial_yf=0$.
The equation along the $y$-direction implies that 
the mass spectrum is characterized by a mass gap $3H/2$ \cite{Langlois:2000ns}.
 The corresponding eigenfunctions 
$f$ can be written in terms of the associated Legendre functions.

Let us now focus on a single KK mode, which is spatially homogeneous
and sufficiently massive: $m\gg H$.
One finds   from (\ref{dS_scalar})
\begin{eqnarray}
\varphi(t)=\frac{1}{ a^{3/2}}\,\cos(mt)\,.
\label{varphi}
\end{eqnarray}
If we take a time average over a time scale much longer than the
period of oscillation $m^{-1}$, we can ignore the oscillatory
behavior and use
\begin{eqnarray}
\bigl\langle\sin^2(mt)\bigr\rangle
=\bigl\langle\cos^2(mt)\bigr\rangle=\frac{1}{2}\,,
\quad\mbox{etc.} \label{timeavr}
\end{eqnarray} 
{}From Eq.~(\ref{Tb}), we thus find 
\begin{eqnarray}
&&  T^{(b)t}{}_{t}
       =
        -\frac{1}{8} |f_m|^2 m^2 \frac{1}{a^3}\,,
\nonumber \\
&& T^{(b)i}{}_{j}
       = 
           \frac{5}{24} |f_m|^2 m^2 \frac{1}{a^3}\delta^{i}_{j}\,,
\end{eqnarray}
where $f_m$ is the value  of $f(y)$ on the brane for the eigenvalue $m^2$.
{}From Eqs.~(\ref{Ett}) and (\ref{varphi}), and from the fact that
$\partial_y^2\phi=-m^2 \phi$ on the brane, 
we can evaluate $E_{\mu\nu}$ as
\begin{eqnarray}
&& - E^{t}{}_{t}
       =\frac{5}{8}\kappa_{5}^2 |f_m|^2 m^2 \frac{1}{a^3}\,,
\nonumber \\
&& -E^{i}{}_{j}
       =-\frac{5}{24}\kappa_{5}^2 |f_m|^2 m^2 \frac{1}{a^3}\delta^{i}_{j}\,,
\end{eqnarray}
where we have neglected the terms that depend on the initial data,
which behave as $a^{-4}$ and thus become negligible at late times.

The above results show that the Weyl term 
$E_{\mu\nu}$ contributes  {\sl negatively} to the 
effective  energy density  and pressure on the brane
for a massive mode.  Moreover, if one computes the 
total effective contribution of the bulk, i.e., the sum of
$T^{(b)}{}_{\alpha\beta}$
and of the Weyl term $E_{\alpha\beta}$, one finds for the effective 
energy density and pressure on the brane
\begin{eqnarray}
&& \kappa_{4}^2 \rho_{({\rm eff})}=-\Bigl(\kappa_{5}^2 {T}
     ^{(b)}{}^{t}{}_{t}-{E}^{t}{}_{t}\Bigr) 
     =
       -\frac{1}{2} \kappa_{5}^2 |f_m|^2 m^2 \frac{1}{a^3}\,,
\nonumber \\
&& \kappa_{4}^2 p_{({\rm eff})}=\frac{1}{3}
\Bigl(\kappa_{5}^2{T} ^{(b)}{}^{i}{}_{i}-{E}^{i}{}_{i}\Bigr) 
     =    0\,. \label{em_dS}
\end{eqnarray}
This represents the backreaction
effects of the bulk scalar field, which are of order ${\cal O}(\phi^2)$.
Whereas the effective pressure due to the  KK mode vanishes, because 
the bulk component and the Weyl component exactly cancel each other, 
the effective energy, remarkably, is negative.

\subsection{KK mode for a low energy cosmological brane}

We now  calculate the effective energy
density and pressure of a KK mode for a low energy cosmological brane. 
The bulk geometry around a brane, with a flat FLRW
geometry and located at $y=0$ is given by the metric  \cite{bdl,bdel}
\begin{eqnarray}
ds^2 = -N^2(t,y) dt^2 
           +Q^2(t,y) a^2(t) d \mbox{\boldmath $x$}^2+dy^2 \,,
\end{eqnarray}
where
\begin{eqnarray}
&& Q(t,y)=  \cosh(y/\ell)-\eta \sinh(|y|/\ell)\,
\nonumber \\
&& N(t,y)= \cosh(y/\ell) -\left(\eta +\frac{\dot\eta}{H}\right)
                               \sinh(|y|/\ell),
\end{eqnarray}
with 
\beq
\eta=\sqrt{H^2\ell^2+1}.
\eeq
We have assumed that there is no dark radiation, i.e., that the bulk 
geometry is strictly AdS and not Schwarzschild-AdS. 
In general, this metric is non-separable. However, in the 
 low energy limit characterized by 
 $H\ell \ll 1$ and $\dot{H}\ell^2\ll 1$, 
we have $\eta\simeq 1$ and $\dot\eta/H\ll1$ so that 
the metric can be approximated by
\beq
ds^2 = dy^2 +e^{-2|y|/\ell}\bigl(-dt^2 +a^2(t)d\mbox{\boldmath $x$}^2
\bigr)\,,
\eeq
which is now separable. If one considers 
 the evolution of a massless, minimally coupled scalar field in  the
above background metric, one finds that the field equation is separable 
and thus admits a solution of the form $\phi(t,y)=f(y)\varphi(t)$ with
\begin{eqnarray} 
&& \partial_y^2 f -\frac{4}{\ell}\partial_y f+m^2 e^{2y/\ell}f=0\,,
\nonumber\\
&& \ddot{\varphi}+3H \dot{\varphi} +m^2\varphi=0\,,
\end{eqnarray}
where the function $f(y)$ is assumed to be $Z_2$-symmetric.

The solution for $f(y)$ with the appropriate Neumann boundary condition
on the brane, $f'(0)=0$ is given in terms of the Hankel functions.
There is a zero mode corresponding to $m=0$ as well as 
a continuum  of KK modes with $m>0$.
For a massive KK mode $m \gg H$, the four-dimensional part evolves as
\begin{eqnarray}
\varphi = \frac{1}{a^{3/2}}\,\cos(mt)\,, 
\end{eqnarray}
Similarly to the de Sitter brane case, one can compute the projection of
the bulk energy-momentum tensor on the brane and one finds for its 
components: 
\begin{eqnarray}
&& T^{(b)}{}^{t}{}_{t}
     =-\frac{1}{4a^3(t)}|f_m|^2 m^2 \bigl\langle\sin^2(mt)\bigr\rangle
     =-\frac{1}{8a^3(t)}|f_m|^2 m^2\,,
\nonumber \\
&& T^{(b)}{}^{i}{}_{i}
      =\frac{5}{4a^3(t)}|f_m|^2 m^2\bigl\langle\sin^2(mt)\bigr\rangle
      =\frac{5}{8a^3(t)}|f_m|^2 m^2\,.
\end{eqnarray}
This gives
\begin{eqnarray}
\kappa_{5}^{-2} E^{t}{}_{t}
       =  \frac{1}{a^4}
       \int^t_{t_0} dt' a^{4}
       \Bigl(
          \partial_{t}   T^{({\rm b})}{}^{t}{}_{t}
       +3\frac{\dot{a}}{a}  T^{({\rm b})}{}^{t}{}_{t}
       -\frac{\dot{a}}{a}T^{({\rm b})}{}^{i}{}_{i}  
       \Bigr) 
 =-\frac{5}{8a^3(t)}|f_m|^2 m^2 \left(1-\frac{a(t_0)}{a(t)}\right)\,,
\end{eqnarray}
and $E^{t}{}_{t}=-E^{i}{}_{i}$.
Thus we obtain 
\begin{eqnarray}
&& T^{(b)}{}^{t}{}_{t}-\kappa_{5}^{-2}E^{t}{}_{t}
   =\frac{1}{2a^3(t)}|f_m|^2 m^2\,,
  \nonumber \\
&&  T^{(b)}{}^{i}{}_{i}-\kappa_{5}^{-2}E^{i}{}_{i}
   =0\,,
\end{eqnarray}
 at late times.
Therefore, the effective energy density and pressure for a KK mode becomes
\begin{eqnarray}
&&\kappa_{4}^2\rho_{\rm (eff)}
  =-\Bigl(\kappa_{5}^2 {T} ^{(b)}{}^{t}{}_{t}-{E}^{t}{}_{t}\Bigr)
  = -\frac{\kappa_{5}^2}{2 a^3(t)} |f_m|^2 m^2 \,, \nonumber  \\
&& \kappa_{4}^2p_{\rm (eff)}
 =\frac{1}{3}
\Bigl(\kappa_{5}^2{T} ^{(b)}{}^{i}{}_{i}-{E}^{i}{}_{i}\Bigr) 
 =0\,.     \label{em_frw}
\end{eqnarray}
This means that, also for a low energy cosmological brane, a massive 
KK mode behaves as cosmic dust with negative energy density.

\vspace{5mm}
The analyses given above imply that the result is independent of 
the existence of a mass gap and the essential factor is
the background expansion of the brane.
  A KK mode can be approximately defined only for a
cosmological brane which slightly deviates from the dS geometry and 
for a low energy brane, thus we expect that our result can be applied at
least for these cases.
 However, for intermediate energy scales a KK mode is not well-defined in
general and it is not clear how our result might be applied.

Finally, we note that the bulk energy density of a KK mode on the
brane remains positive as
\beq
    \kappa_{5}^2 \rho_{({\rm bulk})}
             :=  -\kappa_{5}^2{\cal T}{}^{t}{}_{t}      
              = \frac{1}{4}\kappa_{5}^2
                 |f_m|^2 m^2\frac{1}{a^3}
                >0\,,             \label{bulk_KK}
\eeq       
for both de Sitter and low energy branes
(with the understanding that the time average over scales greater
than $m^{-1}$ is taken).
It shows that there is no singular effect in the bulk in 
contrast to the peculiar behavior on the brane.

\section{Effective theory in the bulk and on the brane including the
gravitational backreaction} 

After having studied the backreaction of the KK modes of a bulk scalar 
field, we now turn to the main subject of this paper, which is to study 
 the backreaction of the gravitational perturbations of the metric itself 
on the cosmology of the brane. 
In this section, we adopt a more general perspective by considering 
a $(d-1)$-brane embedded in a $(d+1)$-dimensional bulk spacetime, although 
we remain primarily interested by the case $d=4$. 
This allows us to investigate the dependence on the number of dimensions 
of various quantities introduced in this section.

\subsection{Effective theory in the bulk}

We now consider only pure gravity in the bulk. 
The action of the system  is given by 
\begin{eqnarray}
 S[g] =\frac{1}{2\kappa_{d+1}^2} 
        \int d^{d+1} x \sqrt{-g}
              \Bigl(
                 {}^{(d+1)}R-2 \Lambda_{d+1}
              \Bigr)
        -\int d^d x \sqrt{-q}\,\sigma   \,,
\end{eqnarray} 
where $\Lambda_{d+1}$ is the bulk cosmological constant and
$\sigma$ is the brane tension.
We mainly consider a dS brane background in this section and
assume that its tension is larger than that of the corresponding RS value 
$2(d-1)/(\kappa_{d+1}^2\ell)$, 
where $\ell=(-d(d-1)/(2\Lambda_{d+1}))^{1/2}$ is the bulk AdS
curvature radius.

We start from an unperturbed metric $\stackrel{(0)}{g}$, which 
is a solution of Einstein's equations and thus satisfies
\begin{eqnarray}
       \frac{\delta S}{\delta g}\bigl[g \bigr]
        \Big|_{\stackrel{(0)}{g}}
     =0\,,
\label{background} 
\end{eqnarray}
where and in what follows the notation,
$Q\bigl[a+g\bigr]\Big|_{f}\,,$
means that a functional $Q[a+g]$ of $g$ is evaluated for
a function $f$, i.e.,
\begin{eqnarray}
Q\bigl[a+g\bigr]\Big|_{f}=Q\bigl[a+f\bigr]\,.
\end{eqnarray}

We then consider (small) linear perturbations of this metric, which we
write $\epsilon \stackrel{(1)}{g}$ and such that its
{\it  average} vanishes 
 i.e., 
\beq
\label{average}
\langle \stackrel{(1)}{g} \rangle =0.
\eeq 
Here we should specify our definition of averaging. 
We assume that the perturbation $\stackrel{(1)}{g}$ has
a typical wavelength $\lambda$ which is much smaller than
the characteristic curvature radius $L$ of the background 
$\stackrel{(0)}{g}$, $\lambda\ll L$. Then we take
the average over a length scale much larger than $\lambda$
but much smaller than $L$. In our case, we can take this
average in the spacetime dimensions parallel to the brane.
However, the situation is dramatically different in the
direction of the extra spatial dimension because the brane
is infinitesimally thin, which implies that the curvature radius
along the extra dimension is infinitely small. Therefore
one cannot take an average in that direction at or around the brane.
Thus our averaging will include only the average over the 
 $1+(d-1)$ spacetime dimensions.
(For spatially homogeneous perturbations, we take only the
time average.)

What we are interested in is the correction to the original metric
due to the backreaction of the metric perturbations.
The total metric we consider can thus be 
written as
\begin{eqnarray}
              g_{\rm tot}=  \stackrel{(0)}{g}
                 +\epsilon \stackrel{(1)}{g}
                 +\epsilon^2 \stackrel{(2)}{g}\,,
\end{eqnarray}
where the quantity $\stackrel{(2)}{g}$ represents the  backreaction
due to the metric perturbations, so that the effective background
(homogeneous) metric, after averaging, is given by 
\begin{eqnarray}
 \bar{g} = \stackrel{(0)}{g}
                 +\epsilon^2 \stackrel{(2)}{g}. 
\end{eqnarray}
For convenience, the parameter $\epsilon$ is introduced as
an expansion parameter, which is to be set to unity at the end of
the calculation.

If we expand the action with respect to $\stackrel{(1)}{g}$,
we have
\begin{eqnarray}
   S\Bigl[\bar{g}+\epsilon \stackrel{(1)}{g}\Bigr]
  =   S\Bigl[\bar{g}\Bigr]
   + \frac{\delta S}{\delta g}\bigl[g \bigr] \Big|_{\bar{g}}
     \Bigl( \epsilon \stackrel{(1)}{g}\Bigr)
   + \frac{1}{2}
     \frac{\delta^2 S}{\delta g^2}\bigl[g \bigr]
     \Big|_{\bar{g}} 
     \Bigl( \epsilon \stackrel{(1)}{g}\Bigr)^2
   +{\cal O}(\epsilon^3)\,.
 \label{expansion}
\end{eqnarray}
Hence the variation of the above expression with respect to 
$\stackrel{(1)}{g}$ yields
\begin{eqnarray}
  \epsilon\frac{\delta S}{\delta g}[\bar{g}+g]
     \Big|_{\epsilon \stackrel{(1)}{g}}
   =  \epsilon  \frac{\delta S}{\delta g}[g] \Big|_{\bar{g}}
     +{\cal O}(\epsilon^2) 
   = \epsilon  \frac{\delta S}{\delta g}[g] \Big|_{\stackrel{(0)}{g}}
     +{\cal O}(\epsilon^2) 
   = {\cal O}(\epsilon^2)\,,
\end{eqnarray}
where we have used Eq.~(\ref{background}) in the final equality.
This implies that, up to ${\cal O}(\epsilon)$,
 the equation of motion for the perturbation
$\stackrel{(1)}{g}$ is given by
\begin{eqnarray}
           \frac{\delta S}{\delta g}
          [\stackrel{(0)}{g}+g]\Big|_{\epsilon \stackrel{(1)}{g}} =0\,.
\end{eqnarray}
On the other hand, the variation of the action with respect to
 $g_{\rm tot}$
 gives 
\begin{eqnarray}
0 = \frac{\delta  S}{\delta g}
            \Bigl[g\Bigr]
            \Bigl|_{ g_{\rm tot}}
 & =&  \frac{\delta  S}{\delta g}
            \Bigl[g\Bigr]
            \Bigl|_{\bar{g}+\epsilon \stackrel{(1)}{g}}
\nonumber \\ 
 &=& \frac{\delta S}{\delta g} 
            \Bigl[g \Bigr] 
            \Bigl|_{\bar{g}}
   + \frac{\delta^2 S}{\delta g^2}\bigl[g \bigr] 
     \Big|_{\bar{g}}
     \Bigl( \epsilon \stackrel{(1)}{g}\Bigr)
   + \frac{1}{2}
     \frac{\delta^3 S}{\delta g^3}\bigl[g \bigr]
     \Big|_{\bar{g}} 
     \Bigl( \epsilon \stackrel{(1)}{g}\Bigr)^2
   +{\cal O}(\epsilon^3)\,
\nonumber \\ 
&=&\frac{\delta S}{\delta g} 
            \Bigl[g \Bigr] 
            \Bigl|_{\bar{g}}
   + \frac{\delta^2 S}{\delta g^2}\bigl[g \bigr] 
     \Big|_{\bar{g}}
     \Bigl( \epsilon \stackrel{(1)}{g}\Bigr)
   + \frac{1}{2}
     \frac{\delta^3 S}{\delta g^3}\bigl[g \bigr]
     \Big|_{\stackrel{(0)}{g}} 
     \Bigl( \epsilon \stackrel{(1)}{g}\Bigr)^2
   +{\cal O}(\epsilon^3)\,,
\end{eqnarray}
where, to get the last expression, 
the argument of the coefficient of the third term, $\bar{g}$,  has
been replaced by $\stackrel{(0)}{g}$, which is
justified within the accuracy of ${\cal O}(\epsilon^2)$.
If one averages the above expression, 
 the second term on the right-hand side vanishes and
we obtain the equation that determines the backreaction-corrected 
background metric $\bar{g}$, in the form
\begin{eqnarray}
 \frac{\delta S}{\delta g} 
            \Bigl[g \Bigr] 
            \Bigl|_{\bar{g}}
 =  -\frac{1}{2}\epsilon^2
      \Big \langle
          \stackrel{(1)}{g}
     \frac{\delta^3 S}{\delta g^3}\bigl[g \bigr]
      \Big|_{\stackrel{(0)}{g}} 
          \stackrel{(1)}{g}
      \Big \rangle   \,. \label{background2}
\end{eqnarray}

Substituting the explicit form for the braneworld 
action, 
we find that 
Eq.~(\ref{background2}) yields 
\begin{eqnarray}
{}^{(d+1)} \bar{G}{}^{a}{}_{b}
          +\Lambda_{d+1} \delta^{a}{}_{b} 
                  =\kappa_{d+1}^2 {\cal T}{}^{a}{}_{b}
                   + \bar{t}_{\rm (brane)}{}^{a}{}_{b} 
                   +\delta {t}_{\rm (brane)}{}^{a}{}_{b}\,,
\end{eqnarray} 
where ${}^{(d+1)} \bar{G}$ is the background bulk 
Einstein tensor including the backreaction effects, i.e., for the metric 
$\bar{g}$. 
And the stress-energy tensor due to the backreaction in the bulk
is given by
\begin{eqnarray}
  \kappa_{d+1}^2 {\cal T}{}^{a}{}_{b}
 =-\Big \langle
        {}^{(d+1)}{\stackrel{(2)}{G}}{}^{a}{}_{b}
   \Big \rangle\,,   \label{bulk_stress}
\end{eqnarray}
where ${}^{(d+1)}{\stackrel{(2)}{G}}{}^{a}{}_{b}$ is the bulk 
Einstein tensor at quadratic order. Here it may be worth noting
that averaging is necessary for this effective stress-energy
tensor to be physically meaningful, since there exists no
locally covariant gravitational energy-momentum tensor
due to the equivalence principle.
The tensor $\bar{t}_{\rm (brane)}{}^a{}_b$ corresponds 
to the brane energy-momentum tensor in the background configuration 
defined by the metric $\bar{g}$ and thus comes from the 
variation of the brane action in the left-hand side of (\ref{background2}).
Finally, $\delta t_{\rm (brane)}{}^a{}_b$, which comes from the 
brane-dependent part 
in the right-hand side of (\ref{background2}),
 denotes the backreaction due to the brane
fluctuations and will be discussed in Section~III.~C.   
The existence of this term is the most important difference 
when compared to the case of the scalar field, in which case
the backreaction originates purely from the bulk.

 Hereafter, we write ${}^{(d+1)} \bar{G}{}^{a}{}_{b}$ as
${}^{(d+1)}G{}^{a}{}_{b}$ for simplicity.
For the moment, we concentrate on the effective theory in the bulk,
\begin{eqnarray}
    {}^{(d+1)}G^{a}{}_{b}
       +\Lambda_{d+1} \delta^{a}{}_{b}
    =\kappa_{d+1}^2 {\cal T}{}^{a}{}_{b}\,.
\end{eqnarray}
Our first task is to evaluate the effective bulk energy-momentum tensor 
${\cal T}^{a}{}_{b}$, which is 
quadratic in the metric  perturbations. Then we will take the limit to
the brane. 

We  now identify the background metric $g^{(0)}$ with the separable metric of  
AdS${}_{d+1}$ bulk-dS brane spacetime and
 $g^{(1)}$ as the linear perturbation of this system. Namely,
\beq
   ds^2 = dy^2 
        + b^2(y)\bigl(\gamma_{\mu\nu}+h_{\mu\nu}
               \bigr)dx^{\mu}dx^{\nu}\,,\quad
     h^{\alpha}{}_{\alpha}  
    =  h_{\alpha}{}^{\beta}{}_{|\beta}
    =0\,,
\eeq
where $b(y)$ is the warp factor defined in Eq.~(\ref{warp_dS}) and
$\gamma_{\mu\nu}$ is the metric of a $d$-dimensional dS
spacetime which is an extension of Eq.~(\ref{dS_four}). 
Note that we have adopted the so-called RS gauge for the
perturbations~\cite{Randall:1999vf,Gen:2000nu}.
The equation of motion for the  perturbations in the bulk reads 
\beq
     \Bigl[
        \frac{1}{b^d}\partial_{y} \Bigl(b^d \partial_{y} \Bigr)
         +\frac{1}{b^2}\Bigl( \Box_{d}-2H^2 \Bigr)
     \Bigr]h_{\alpha\beta}=0\,. \label{eom}
\eeq
This equation is separable and one considers solutions of the form
$h_{\alpha\beta}= f(y)\varphi_{\alpha\beta}(x^{\mu})$, 
where $f(y)$ is the generalization of the solution of Eq.~(\ref{dS_scalar})
to the case of a $d$-dimensional brane with boundary
condition $\partial_y f(y)=0$ at $y=y_0$
because $\partial_y h_{\alpha\beta}=0$ on the brane.
Similarly  to the scalar case, the separation constant $m$
represents the effective mass of a KK graviton mode  and satisfies 
$m>(d-1)H/2 $. The $d$-dimensional part
$\varphi^{\alpha}{}_{\beta}$ satisfies
\beq
 \Bigl[ \Box_{d}-2H^2\Bigr]\varphi^{\alpha}{}_{\beta}=
  m^2\varphi^{\alpha}{}_{\beta}\,.
\eeq
 We focus on a KK mode with $m^2 \gg H^2$.
Furthermore, for simplicity, we focus on perturbations of the
tensor-type with respect to the spatial $(d-1)$-geometry, namely
on those with $h^{t}{}_{t}=h^{t}{}_{i}=h^{i}{}_{t}=0$.
Taking the slicing of the de Sitter space with 
the flat spatial $(d-1)$-geometry, they will have the form,
\beq
h^{i}{}_{j} =\frac{f_m}{a^{(d-1)/2}}\, \cos(mt)\, Q^{i}{}_{j}\,,
\label{tensor_time}
\eeq
where $f_m$ is the amplitude of the KK mode 
 and $Q^{i}{}_{j}$ is the polarization tensor on the flat $(d-1)$-space.
The amplitude $f_m$ can be determined, for instance, by the normalization
condition if one considers a quantized perturbation theory.

As mentioned earlier, in order to obtain the stress-energy tensor
that embodies the backreaction due to the metric perturbations, one
needs to ``average''  the Einstein tensor at quadratic order, according to  
Eq.~(\ref{bulk_stress}).
The components of the bulk curvature tensors, up
to quadratic order in the perturbations are listed in Appendix~A.
As explained after Eq.~(\ref{average}), we take the spacetime average 
in the $1+(d-1)$ dimensions parallel to the brane, but not
along the extra dimension. In particular, because of the cosmological
symmetry, we can take the average in the $(d-1)$ dimensions over
the complete space.
The derivatives along
the extra dimension are replaced by using the field equation~(\ref{eom})
 and the boundary conditions on the brane. 
Our procedure is detailed in Appendix~B.

Using Eq.~(\ref{bulkcurve}) of Appendix~A and the computational
rules detailed in Appendix~B, we obtain in the limit $y\to+0$ the expressions
\begin{eqnarray}
&&  \Big \langle
    {}^{(d+1)}{\stackrel{(2)}{G}}{}^{y}{}_{y}
    \Big \rangle
=  -\frac{1}{8}\Big \langle
         h^{\rho\sigma}
         \Box_{d}\,  
         h_{\rho\sigma}   
       \Big \rangle  \,,
 \nonumber \\
&&
     \Big \langle
     {}^{(d+1)}\stackrel{(2)}{G}{}^{\alpha}{}_{\beta}
     \Big \rangle
  =  
     -\frac{1}{2}
          \Big \langle
           h^{\alpha\rho}\Box_d h_{\rho\beta}
          \Big \rangle
     -\frac{d-3}{8d} \delta^{\alpha}{}_{\beta}  
          \Big \langle
           h^{\rho\sigma}\Box_d  h_{\rho\sigma}
          \Big \rangle
     -\frac{1}{4}
          \Big \langle
           h^{\rho\sigma|\alpha}h_{\rho\sigma|\beta}
          \Big \rangle  
         \,. 
\label{bulk eff}
\end{eqnarray}
A priori, the effective energy-momentum tensor includes an anisotropic 
stress, to which each mode will contribute with a factor $O(m^2)$.
However, if the perturbations are described by 
a random field which is statistically homogeneous and isotropic,
the average over all modes of the anisotropic part must cancel. What 
remains is thus to justify the randomness of
the perturbations. In this respect, the quantum fluctuations 
are indeed expected to have this property.
Also, $\langle {}^{(d+1)}{\stackrel{(2)}{G}}{}^y{}_\nu \rangle$
vanishes on the brane by using the boundary conditions
 $\partial_y h_{\alpha\beta}=0$ on it.

\subsection{Backreaction on the brane}

Let us now discuss the effect of the backreaction onto the brane.
The projected gravitational equation on the brane reads
\begin{eqnarray}
{}^{(d)}G^{\alpha}{}_{\beta}
           = -\Lambda_{\rm eff}\,  \delta^{\alpha}{}_{\beta}
           +\kappa_{d}^2\, \tau^{\alpha}{}_{\beta}[h,h]
           +\kappa_{d+1}^2\, T^{({\rm b})}{}^{\alpha}{}_{\beta}   
           -E^{\alpha}{}_{\beta}, \label{brane_eff}
\end{eqnarray}
 where 
\beq
   \Lambda_{\rm eff}
    =\frac{d-2}{d}\Lambda_{d+1} 
    +\frac{d-2}{8(d-1)} 
     \kappa_{d+1}^4\sigma^2\,,
\eeq
is the effective cosmological constant on the brane,
and 
\begin{eqnarray}
\kappa_{d+1}^2 T^{({\rm b})}{}^{\alpha}{}_{\beta}
&=&\frac{d-2}{d-1}\kappa_{d+1}^2 
 \Bigl[
    {\cal T}{}^{\alpha}{}_{\beta}
   +\delta^{\alpha}{}_{\beta}
    \Bigl(
         {\cal T}^y_y
      -\frac{1}{d}{\cal T}{}^a{}_{a}
    \Bigr)
 \Bigr]  
\nonumber\\
 &=&-\frac{d-2}{d-1} 
   \Bigl[
        \Big \langle  
       {}^{(d+1)}  \stackrel{(2)}{G}{}^{\alpha}{}_{\beta}
        \Big \rangle
   +\delta^{\alpha}{}_{\beta}
    \Bigl(
        \Big \langle
        {}^{(d+1)} \stackrel{(2)}{G}{}^{y}{}_{y}
        \Big \rangle
     -\frac{1}{d}
        \Big \langle
       {}^{(d+1)} \stackrel{(2)}{G}{}^{a}{}_{a}
        \Big \rangle 
   \Bigr)
 \Bigr]
\nonumber \\
 &=&  
   \frac{d-2}{2(d-1)}
       \Big \langle
          h^{\alpha\rho }\Box_d \,h_{\rho\beta}
       \Big \rangle
   +\frac{(d-2)(d-3)}{8d(d-1)}\delta^{\alpha}{}_{\beta}
         \Big \langle
            h^{\rho\sigma}\Box_d\, h_{\rho\sigma}
         \Big \rangle
   + \frac{d-2}{4(d-1)}
         \Big \langle
           h^{\rho\sigma|\alpha}h_{\rho\sigma|\beta}
         \Big \rangle\,,
\end{eqnarray}
is the projection of the effective energy-momentum tensor of the
bulk gravitons.
The tensor $\tau^{\alpha}{}_{\beta}$, corresponding to 
$\delta t_{\rm (brane)}{}^{a}{}_{b}$  
of the previous subsection, describes the brane perturbation induced by
the bulk perturbation.
 We will show in the next subsection that, for our purposes,
this term can be neglected. 
We now concentrate on the effect of the effective 
energy-momentum of the bulk gravitons projected
on the brane, i.e., the terms $T^{(b)\alpha}{}_\beta$ and $E^\alpha{}_\beta$.

Let us first consider $T^{(b)\alpha}{}_\beta$.
Because of the assumed symmetries, i.e., the spatial homogeneity
and isotropy, this gives in the brane an effective
perfect fluid with some energy density and  pressure.
Decomposing the metric perturbations into KK modes,
one finds that the contribution of a sufficiently massive mode
to the energy density and pressure is given by 
\begin{eqnarray}
&&  \kappa_{d+1}^2 T^{({\rm  b})}{}^{t}{}_{t}
 =   -\frac{(d+3)(d-2)}{16d(d-1)}
    \frac{1}{a^{d-1}}
     m^2|f_m|^2
    \Big \langle
          Q^{k\ell}Q^{\ast}_{k\ell}
    \Big \rangle  
\,,
\nonumber \\
&&  \kappa_{d+1}^2 T^{({\rm  b})}{}^{i}{}_{j}
 = \frac{(d^2+3)(d-2)}{16d (d-1)^2}
       \frac{1}{a^{d-1}}
     m^2 |f_m|^2
     \Big \langle
      Q^{k\ell}Q^{\ast}_{k\ell}    
     \Big \rangle \,
     \delta^{i}{}_{j}\,.
\end{eqnarray}

We must also take into account the projection of the 
Weyl tensor on the brane,  $E^\alpha{}_\beta$.
Although this term is not included in the "bulk
energy-monemtum" tensor because it is 
a part of the
bulk Weyl tensor, it contributes nevertheless to the projected
gravitational equations 
as an ``energy-momentum" tensor.
Although its direct evaluation is rather delicate, this term can be 
computed by resorting once more to the cosmological symmetry.
{}From Eq.~(\ref{brane_eff}), the contracted Bianchi identities
$D^{\alpha}{}^{(d)}G_{\alpha}{}^{\beta}=0$, together with the
conservation of 
$\tau_{\alpha}{}^{\beta}$, give 
\begin{eqnarray}
D^{\mu}E_{\mu\nu}
    = \kappa_{d+1}^2  D^{\mu} T^{({\rm b})}_{\mu\nu}.   
\end{eqnarray}
Because of the cosmological symmetry, the only non-trivial
component of the above equation is the time component, which reads 
\beq
\partial_t  E^{t}{}_{t}
+d\frac{\dot{a}}{a}  E{}^{t}{}_{t}=\kappa_{d+1}^2
\Bigl(
\partial_{t}   T^{({\rm b})}{}^{t}{}_{t}
       +(d-1)\frac{\dot{a}}{a}  T^{({\rm b})}{}^{t}{}_{t}
       -\frac{\dot{a}}{a}T^{({\rm b})}{}^{i}{}_{i}
\Bigr)\,,
\eeq
where, on the left-hand side, we have used the property that $E_{\mu\nu}$ is 
traceless and thus $E^i{}_i=-E^{t}{}_{t}$. The integration then yields
\begin{eqnarray}
 E^{t}{}_{t}
      =  \frac{\kappa_{d+1}^2}{a^d}
       \int^t_{t_0} dt' a^{d}
       \Bigl(
          \partial_{t}   T^{({\rm b})}{}^{t}{}_{t}
       +(d-1)\frac{\dot{a}}{a}  T^{({\rm b})}{}^{t}{}_{t}
       -\frac{\dot{a}}{a}T^{({\rm b})}{}^{i}{}_{i}  
       \Bigr)  \,. \label{ett}
\end{eqnarray}
As before, we neglect the contribution from the initial condition,
which is valid at late times.

Substituting a KK graviton mode given by Eq.~(\ref{tensor_time})
into the integrand on the right-hand side of Eq.~(\ref{ett}), 
and taking the time average, one finds
\begin{eqnarray}
\kappa_{d+1}^2
        \Bigl(
          \partial_{t}   T^{({\rm b})}{}^{t}{}_{t}
       +(d-1)\frac{\dot{a}}{a}  T^{({\rm b})}{}^{t}{}_{t}
       -\frac{\dot{a}}{a}T^{({\rm b})}{}^{i}{}_{i}  
        \Bigr)
= -\frac{(d^2+3)(d-2)}{16d(d-1)}\,\frac{H}{a^{d-1}}\,|f_m|^2 m^2
    \Big \langle Q^{k\ell}Q_{k\ell}^*\Big \rangle\,.
\end{eqnarray}
This gives, at late times,
\begin{eqnarray}
{E}^{t}{}_{t}
  = -\frac{(d^2+3)(d-2)}{16d(d-1)}\,
    \frac{1}{a^{d-1}}\,m^2 |f_m|^2
\Big \langle Q^{k\ell}Q_{k\ell}^*\Big \rangle.
\end{eqnarray}
Because of the traceless nature of this tensor, we then obtain
${E}^{i}{}_{j}=-(1/(d-1)){E}^{t}{}_{t}\delta^i_j$. 

The total contribution of the two tensors is therefore 
\beq
\kappa_{d+1}^2{T}^{({\rm b})}{}^{t}{}_{t}
  - {E}{}^{t}{}_{t}
    = \frac{d-2}{16}\,\frac{1}{a^{d-1}}\, m^2 |f_m|^2
          \Big \langle
         Q^{k\ell}Q^{\ast}_{k\ell}
          \Big \rangle 
         \,.
\eeq
for the temporal part and 
\beq
\kappa_{d+1}^2{T}^{({\rm b})}{}^{i}{}_{i}- {E}{}^{i}{}_{i} = 0 ,
\eeq
for the spatial part.
This means that the contributions of a KK mode to the total effective
energy density and pressure are respectively given by 
\begin{eqnarray}
&& \kappa_{d}^2 \rho_{({\rm eff})}
   = -\frac{d-2}{16}\,\frac{1}{a^{d-1}}\, m^2 |f_m|^2
          \Big \langle
         Q^{k\ell}Q^{\ast}_{k\ell}
          \Big \rangle 
         \,,
\nonumber \\
&& \kappa_{d}^2 p_{({\rm eff})}
   = 0\,.
\end{eqnarray}
For instance, for $d=4$, we obtain
\begin{eqnarray}
&& \kappa_{4}^2 \rho_{({\rm eff})}
   = -\frac{1}{8\,a^3} m^2 |f_m|^2
          \Big \langle
         Q^{k\ell}Q^{\ast}_{k\ell}
          \Big \rangle 
         \,,
\nonumber \\
&& \kappa_{4}^2 p_{({\rm eff})}
   = 0\,.
\end{eqnarray}
The effective isotropic pressure vanishes and the
effective energy density is negative.
This is the same as in the case of the scalar field discussed in the
previous section. 

We note that the bulk energy density of a KK mode on the brane remains 
positive
\beq
 \kappa_{d+1}^2 \rho_{({\rm bulk})}
:= - \kappa_{d+1}^2 {\cal T}{}^{t}{}_{t}
 = \frac{d+3}{16d}\,\frac{1}{a^{d-1}}\,m^2|f_m|^2
           \Big \langle
                Q^{k \ell}Q^{\ast}_{k\ell}
           \Big \rangle
         >0\,,
\eeq
as in the scalar case, Eq.~(\ref{bulk_KK}).
It shows again that there is no singular effect in the bulk.
The negativity of the effective energy density on the brane originates
from the projected Weyl tensor $E_{\mu\nu}$.

\subsection{Brane intrinsic contributions}

We now consider the brane intrinsic contributions.
In order to discuss the gravitational perturbations in the brane
world, it is  not sufficient to consider the contribution from the
bulk.  The brane perturbations must be taken into account as well. 
We take an approach in which we derive the second order boundary
action and regard it as the action for an effective matter on the
brane.

In this paper, the brane is treated as  a thin wall.
In the thin wall approximation, the second order action on the
boundary has been derived in the Appendix of~\cite{Gen:2000nu}.
When there is no ordinary matter on the brane and thus no brane
bending mode, the second order boundary action is
given by~\cite{Gen:2000nu}
\begin{eqnarray}
\delta^{2} S
 =\frac{1}{2\kappa_{d+1}^2}
   \int_{\partial M} d^d x \sqrt{-\tilde{q}}
      \Bigl[
          -\Delta k^{\rho\sigma}\tilde{h}_{\rho\sigma}
          +\frac{\sigma\kappa_{d+1}^2}{2(d-1)}
              \tilde{h}^{\rho\sigma}\tilde{h}_{\rho\sigma}
      \Bigr]\,,
\end{eqnarray}
where $\tilde{h}_{\mu\nu}=b^2 h_{\mu\nu}$,
$k_{\rho\sigma}=\partial_y\tilde{h}_{\rho\sigma}/2$ 
and $\Delta Q=Q^{(+)}-Q^{(-)}$.
For an  AdS-bulk configuration and with the assumption of $Z_{2}$ symmetry
about the brane, this reduces to
\begin{eqnarray}
\delta^{2} S
   = \frac{3}{4(d-1)} \sigma 
       \int_{\partial M}
          d^d x \sqrt{-\tilde{q}}\, h_{\rho\sigma}h^{\rho\sigma}\,.
\end{eqnarray}
The second order action can be regarded as an action for some effective matter
induced on the brane 
\begin{eqnarray}
      \int_{\partial M}   d^d x \sqrt{-\tilde{q}}\, {\cal L}_{m}\,,\quad
     {\cal L}_{m}:=\frac{3}{2(d-1)} \sigma
                   h_{\rho\sigma}h^{\rho\sigma}\,. 
\end{eqnarray}
Its variation with respect to the background metric $\tilde{q}_{\mu\nu}$
yields the induced matter energy-momentum tensor on the brane
\begin{eqnarray}
&&  \tau_{\alpha\beta}
    =\frac{2}{\sqrt{-\tilde{q}}}
       \frac{\delta}{\delta \tilde{q}^{\alpha\beta}}
        \Bigl(
          \sqrt{-\tilde{q}}\,{\cal L}_{m}
        \Bigr)
= \frac{6}{d-1}\sigma 
   \Bigl(h_{\alpha\rho} h^{\rho}{}_{\beta}
        -\frac{1}{4}\tilde{q}_{\alpha\beta}h^{\rho\sigma}h_{\rho\sigma}        
   \Bigr)
     \,.
\end{eqnarray}
Note that, strictly speaking,  $\tilde{q}_{\mu\nu}$ does not include
the backreaction. However, as discussed in Section~III.~A, for 
perturbations with small amplitude, the linear perturbation equations 
are identical to those for the background metric in which the 
backreaction is taken into account.
Thus we can add this term as a part of the (effective)
matter contribution in the effective equation on the brane.

We can readily calculate the effective energy density and pressure
of this contribution. One finds
\begin{eqnarray}
&& \kappa_{d}^2\rho_{\rm (brane)}
  =\frac{3(d-2)}{16(d-1)^2} \kappa_{d+1}^4\sigma^2
    |f_m|^2\frac{1}{a^{d-1}}
   \Big \langle
     Q^{k\ell} Q^{\ast}_{k\ell}
   \Big \rangle \,,
\nonumber  \\
&& \kappa_{d}^2 p_{\rm (brane)}
 =\frac{3(d-2)(-d+5)}{16(d-1)^3} \kappa_{d+1}^4\sigma^2
    |f_m|^2\frac{1}{a^{d-1}}
   \Big \langle
     Q^{k\ell} Q^{\ast}_{k\ell}
   \Big \rangle \,.
\end{eqnarray}
This gives the equation of state 
\begin{eqnarray}
w_{\rm (brane)}=- \frac{d-5}{d-1}\,.
\end{eqnarray}
For  $d=4$ the boundary contribution thus behaves as  radiation,
for $d=5$ as  dust and for $d>6$ as  matter with negative pressure.
However,  its contribution to ${}^{(d)}G^{\alpha}{}_{\beta}$ is of order
\begin{eqnarray}
   \kappa_{d+1}^4 \sigma \tau^{\alpha}{}_{\beta} 
   \sim \frac{1}{\ell^2}
      \Bigl(
         1+\bigl(H\ell \bigr)^2
      \Bigr) 
   h^{\rho\sigma}h_{\rho\sigma}\,, \label{brane fluc}
\end{eqnarray}   
where we use~\cite{Gen:2000nu}
\beq
 \sigma = \frac{2(d-1)}{\kappa_{d+1}^2\ell}
            \Bigl(1 +\bigl(H\ell\bigr)^2\Bigr)
            ^{1/2}
  \,.
\eeq
For the cases $H\ell\ll 1$ and $H\ell \gg 1$, the right-hand side of
Eq.~(\ref{brane fluc}) is ${\cal O}(\ell^{-2}h^2)$ and ${\cal O}(H^2 h^2)$,
respectively.
Thus as long as we consider sufficiently massive KK modes, with
$m\gg {\rm max} \{\ell^{-1}, H\}$, the brane perturbations can be
safely neglected and only the projected bulk contributions are relevant for
the effective theory on the brane.

\section{Negative energy density from the bulk point of view}

Intuitively, the ``negative energy density'' of a KK mode is rather 
puzzling. However, we can understand its cause by regarding the KK
modes as a part of the dark component, such as the dark radiation.
The energy density of the dark component evolves as \cite{Langlois:2003zb}
\begin{eqnarray}
\dot{\rho}_{(D)} +4H \rho_{(D)}= 
       -2\Bigl(1+\frac{\rho}{\sigma}\Bigr){\cal T}_{ab} u^{a}n^{b}
       -2(H\ell) {\cal T}_{ab}n^a n^b\,, \label{dark}
\end{eqnarray} 
where $u^{a}$ and $n^{a}$ are the tangent and normal vectors to the brane,
respectively.
Since there is 
no matter on the brane and no brane-bulk energy
exchange, the first term on the right-hand side of Eq.~(\ref{dark})
vanishes and only the second one, related to the pressure transverse to
the brane, survives. In terms of the energy conservation law in the bulk,
this has the simple interpretation that the work done by the pressure
on the brane to move it outward in the direction of the AdS infinity
reduces the energy in the bulk. As a result, 
the dark energy density decreases, since the dark energy density
on the brane is proportional to the total mass (energy) in the bulk
\cite{Minamitsuji:2003at}.
For a massless scalar field,
\begin{eqnarray}
{\cal T}_{ab}n^a n^b= \frac{1}{2}\dot{\phi}^2>0\,. 
\end{eqnarray}
Thus, the dark component decays faster than ordinary  radiation.
For the KK modes, after time averaging, we have
\begin{eqnarray}
{\cal T}_{ab}n^a n^b= \frac{1}{4\,a^3}\, |f_m|^2 m^2 \,.
\end{eqnarray}

The formal solution of Eq.~(\ref{dark}) is 
\begin{eqnarray}
\rho_{(D)}= -\frac{2}{a^4}\int^{t}_{t_0}
            dt \, a^4  (H\ell) {\cal T}_{ab}n^{a}n^{b}
          +\frac{C}{a^4},
\end{eqnarray}
where $C$ denotes the initial mass in the bulk.
For the KK modes,
\begin{eqnarray}
\int^{t}_{t_0}dt \, a^4 (H\ell) {\cal T}_{ab}n^{a}n^{b}
=\frac{\ell}{4} \,|f_m|^2 m^2\int^{t}_{t_0}dt\, \dot{a}
=\frac{\ell}{4}\,\Bigl(a(t)-a(t_0)\Bigr)|f_m|^2 m^2\,.
\label{int}
\end{eqnarray}
Hence
\begin{eqnarray}
\rho_{(D)}=-\frac{\ell}{2a^3}|f_m|^2m^2+\frac{C}{a^4}\,,
\end{eqnarray}
where we have redefined the mass parameter $C$ by absorbing into it
the initial data dependent term of the integral (\ref{int}).

 Anyway, in the case of a dS brane (or a cosmological brane which
slightly deviates from the dS geometry), the effective cosmological
constant dominates the
cosmological evolution and the KK effect does not have a significant
impact on the brane.
For a low energy brane, especially for a radiation-dominated
brane, naively one might worry that this result would imply
the appearance of a negative energy density within
a finite time.
However from Eq.~(\ref{dark}), the bulk pressure term is 
proportional to $H$. Hence if $\dot H<0$ at $H=0$, the
energy density will remain positive at the expense of
rendering the universe to recollapse.

 For simplicity, we consider the case where the cosmological
evolution of the brane is determined solely by the dark component. 
Note that this  discussion can be generalized  when one considers 
 ordinary dust or radiation in addition to  the dark component.
The Hubble parameter on the brane is obtained from
\begin{eqnarray}
 H^2 =\frac{\kappa_{5}^2}{3 \ell} \rho_{(D)}
     = -\frac{\kappa_{5}^2}{6}\,\frac{1}{a^3}|f_m|^2m^2
       +\frac{\kappa_{5}^2}{3\,\ell}\,\frac{C}{a^4}
     =-\frac{2\kappa_5^2}{3}{\cal T}_{ab}n^an^b
+\frac{\kappa_{5}^2}{3\,\ell}\,\frac{C}{a^4}\,.
\end{eqnarray}
Taking the time derivative of this equation, 
we obtain
\begin{eqnarray}
\dot{H}=\frac{\kappa_{5}^2}{4}\,\frac{1}{a^3}|f_m|^2m^2
            -\frac{2\kappa_{5}^2 }{3\,\ell}\,\frac{C}{a^4}
 =\kappa_5^2\,{\cal T}_{ab}n^an^b 
-\frac{2\kappa_{5}^2 }{3\,\ell}\,\frac{C}{a^4}\,.
\end{eqnarray}
Therefore, at $H=0$, we have
\begin{eqnarray}
\dot{H}= -\frac{1}{3} {\cal T}_{ab} n^a n^b <0\,,
\end{eqnarray}
and the universe begins to collapse.
Thus, the backreaction of the KK modes leads to
a collapsing universe.

The situation is the same for the case of KK gravitons
as long as the brane fluctuations are negligible,
because we have
\begin{eqnarray}
\kappa_{5}^2 \,{\cal T}_{ab}n^a n^b
 = -\Big \langle 
        \stackrel{(2)}{G}{}^{y}{}_{y}
    \Big \rangle
  =\frac{1}{8}m^2
    \Big \langle 
        h^{\rho\sigma}h_{\rho\sigma}
    \Big \rangle
  =\frac{1}{16}|f_m|^2 m^2 \frac{1}{a^3}
         \Big\langle
             Q^{k\ell}Q^{\ast}_{k\ell}
         \Big \rangle>0\,.
\end{eqnarray}
Thus, provided that the brane fluctuations can be neglected, 
the brane universe will start to collapse within a finite time.
For more realistic situations in cosmology, our result
suggests that for a low energy brane the brane universe will
eventually collapse unless the contribution of the true (normal) dust matter is
 larger than that of KK modes.

\section{Summary and discussion}

We have investigated the effect of a Kaluza-Klein (KK) mode 
on  brane cosmology, focusing on the second
Randall-Sundrum (RSII) type model, i.e., for
a $Z_2$-symmetric brane embedded in the anti-de Sitter bulk.

The KK gravitons, which are just the metric perturbations
in the bulk, are produced during a de Sitter (dS) brane inflation
phase via vacuum fluctuations. 
{}From the four-dimensional point of view they are effectively
equivalent to massive gravitons with masses $m>3H/2$, where
$H$ represents the dS expansion rate of the brane.  
The theory of linear perturbations reveals that the squared amplitude of a KK
mode decays as $a^{-3}$ and its contribution rapidly becomes
negligible during the brane inflation.
However, after brane inflation in the radiation-dominated
Friedmann-Lema\^\i tre-Robertson-Walker (FLRW) era the background radiation energy
density decays as $a^{-4}$, which implies that the contribution of KK
gravitons may 
have a significant impact on the brane cosmology at late times.

Before discussing the case of the KK gravitons, we have considered
the case of a massless, minimally coupled bulk scalar field as an exercise.
This is because a massless, minimally coupled scalar field has many properties
in common with the gravitational perturbation, 
but it is much simpler to deal with the former than with the latter.
We have considered two limiting cases of the background spacetime
in which the equations of motion become separable, namely, the case of
a de Sitter brane and the case of a low energy cosmological brane.
In both cases, we found that a sufficiently massive KK mode, with
mass much greater than the Hubble expansion rate, $m\gg H$, 
indeed behaves like dust but, rather surprisingly, its energy
density is negative.

Then we have turned to the case of the bulk gravitational perturbations.
We have first derived the effective energy-monemtum
tensor of a KK mode in the bulk, and investigated the effect
of the KK mode on the brane cosmology by projecting the effective
 energy-monemtum tensor on the brane.
We have found exactly the same behavior as that of the scalar,
i.e., a massive KK mode behaves as a negative energy density dust.

The negative energy density of a KK mode may sound rather puzzling.
But, from the bulk point of view,
we have shown that this result can be regarded as a natural
consequence of the energy conservation law in the bulk.
Here the essence is to recall that
the so-called dark radiation term, which behaves like
radiation on the brane, describes the total mass in the bulk.
Then,  a very massive KK mode corresponds to a particle
with a high momentum in the direction of the extra dimension, which 
exerts a pressure on the brane and pushes it outward in the
 direction of the AdS infinity. As a result, the energy
in the bulk decreases, leading to the decrease of the dark
energy term. Thus, a massive KK mode gives a negative contribution
to the dark radiation term. This is why a KK mode behaves like
a negative energy density dust.

Note that the negative energy of a KK mode emerges only
from the effective four-dimensional point of view on the brane.
The bulk energy density for a KK mode still remains positive
and thus there is no singular effect in the bulk.

We have studied the two cases (de Sitter and low energy branes)
in which the bulk equations are separable, hence the KK modes are
well defined. However, for a general cosmological brane, 
one cannot define a KK mode since its very definition depends on
the separability of the equations in the bulk.
Nevertheless, considering the discussion from the bulk point
of view given in the previous paragraph,
it seems reasonable to expect that this 
backreaction effect of the bulk metric perturbations
persists for a general cosmological brane. 
Thus we conclude that the effect of very massive KK modes
is to reduce the energy density on the brane, hence
the expansion rate, and for a low energy brane the
universe will
recollapse unless the contribution of a normal (true) dust matter 
is larger than that of the KK modes.
 
To quantify this effect in realistic cosmological models,
there are some additional  issues that remain to  be resolved.
In this paper, we have considered only a single KK mode and calculated
the effective energy density and pressure.
In reality, one should  integrate over all the KK modes that contribute to
 the cosmology of the brane. This requires first the knowledge of  the whole 
spectrum of the KK modes, which will presumably be determined by vacuum
fluctuations in the bulk  \cite{Kobayashi:2000yh, Sago:2001gi,
Naylor:2004ua}. However, knowing the whole spectrum 
may not be enough, because a naive integral of the KK spectrum 
is expected to diverge. One 
 would then need an appropriate regularization scheme.
In connection with this, it may be important to take into account
the thickness of a brane, either classically as in the case of a 
classical domain wall or quantum mechanically by considering the
wavefunction of a brane.

\section*{Acknowledgments}
This work was supported in part by Monbu-Kagakusho Grant-in-Aid for
Scienfitic Research (S) No. 14102004.

\appendix

\section{Second order curvature tensors in the bulk}

Here we spell out the components of the curvature tensors
up to quadratic order in the bulk metric perturbation.
We consider the $(d+1)$-dimensional perturbed metric in the form,
\begin{eqnarray}
      ds^2 =dy^2 
           +b^2(y)\Bigl(
                      \gamma_{\mu\nu}+h_{\mu\nu}
                  \Bigr)
                   dx^{\mu}dx^{\nu}\,,
\end{eqnarray} 
where $\gamma_{\mu\nu}$ is the metric of the background $d$-dimensional
spacetime section. In the text, we identify $\gamma_{\mu\nu}$ with 
the metric of a de Sitter spacetime.
We impose the following gauge conditions on the perturbation:
\begin{eqnarray}
  h^{\alpha}{}_{\alpha}= h_{\alpha}{}^{\beta}{}_{|\beta}=0\,,
  \label{conditions}
\end{eqnarray}
where the vertical bar ($\,|\,$) denotes the covariant derivative
 associated with the $d$-dimensional metric $\gamma_{\mu\nu}$,
and the tensor indices of $h_{\mu\nu}$ are raised or lowered by
the metric $\gamma_{\mu\nu}$ (not by the 5-dimensional metric).

The non-trivial components of the connection are given by
\begin{eqnarray}
&&{}^{(d+1)}\Gamma^{\mu}_{y\nu}
   =\frac{b'}{b}\delta^{\mu}{}_{\nu}
   +\frac{1}{2}h^{\mu}{}_{\nu}{}'
     -\frac{1}{2}h^{\mu\alpha}h'_{\alpha\nu}\,,
 \nonumber \\
&&{}^{(d+1)}\Gamma^{y}_{\mu\nu}
   =-b\,b' \gamma_{\mu\nu}-\frac{1}{2}b^2 h'_{\mu\nu}
    -b\,b'h_{\mu\nu}
\nonumber \\
&&{}^{(d+1)}\Gamma^{\mu}_{\alpha\beta}
  =
{}^{(d)}{\Gamma}{}^{\mu}{}_{\alpha\beta}
+\frac{1}{2}\gamma^{\mu\rho}
             \Bigl(h_{\alpha\rho|\beta}+h_{\beta\rho|\alpha}
 -h_{\alpha\beta|\rho} \Bigr)
-\frac{1}{2}h^{\mu\rho}
             \Bigl(h_{\alpha\rho|\beta}+h_{\beta\rho|\alpha}
 -h_{\alpha\beta|\rho} \Bigr),
\end{eqnarray}
where the prime (${~}'$) denotes the $y$-derivative.

The non-trivial components of the Riemann tensor are given by
\begin{eqnarray}
&& {}^{(d+1)} R^{\mu}{}_{y\nu  y}
  = -\frac{b''}{b}\delta^{\mu}{}_{\nu}
    -\frac{1}{2}h^{\mu}{}_{\nu}{}''
    -\frac{b'}{b}h^{\mu}{}_{\nu}{}'  
    +\frac{1}{4}h^{\mu\alpha}{}'h_{\alpha\nu}'
    +\frac{b'}{b}h^{\mu\rho}h_{\rho\nu}'
    +\frac{1}{2}h^{\mu\alpha}h_{\alpha\nu}'' \,,
 \nonumber \\
&&{}^{(d+1)} R^{y}{}_{\mu  y \nu}
= -b\,b''\gamma_{\mu\nu}
   -b\,b''h_{\mu\nu}
   -\frac{1}{2}b^2 h_{\mu\nu}''
   -b\,b'h_{\mu\nu}'
    +\frac{1}{4}b^2 h_{\alpha\nu}'h^{\alpha}{}_{\mu}{}'\,, 
 \nonumber  \\
&& {}^{(d+1)}R^{\alpha}{}_{y\mu\nu}
=  \Bigl(\frac{1}{2}h^{\alpha}{}_{\nu}{}'
       -\frac{1}{2}h^{\alpha\beta}h_{\beta\nu}' \Bigr)_{|\mu}
-  \Bigl(\frac{1}{2}h^{\alpha}{}_{\mu}{}'
       -\frac{1}{2}h^{\alpha\beta}h_{\beta\mu}' \Bigr)_{|\nu}
\nonumber \\
&&  \hspace{1cm}
+\frac{1}{4}h^{\rho}{}_{\nu}{}'
\Bigl(h^{\alpha}{}_{\rho|\mu}+h^{\alpha}{}_{\mu|\rho}
            -h_{\mu\rho}{}^{|\alpha} \Bigr)
-\frac{1}{4}h^{\rho}{}_{\mu}{}'
\Bigl(h^{\alpha}{}_{\rho|\nu}+h^{\alpha}{}_{\nu|\rho}
            -h_{\nu\rho}{}^{|\alpha}  \Bigr)\,,  
 \nonumber \\
&& {}^{(d+1)}R^{y}{}_{\alpha\mu\nu}
    =-\frac{1}{2}b^2\Bigl(h_{\alpha\nu|\mu}
                          -h_{\alpha\mu|\nu} \Bigr)'
-\frac{b^2}{4}h_{\rho\mu}'
     \Bigl(h^{\rho}{}_{\alpha|\nu}+h^{\rho}{}_{\nu|\alpha}
          -h_{\alpha\nu}{}^{|\rho} \Bigr)
 \nonumber \\
&& \hspace{1cm}
+\frac{b^2}{4}h_{\rho\nu}'
     \Bigl(h^{\rho}{}_{\alpha|\mu}+h^{\rho}{}_{\mu|\alpha}
          -h_{\alpha\mu}{}^{|\rho} \Bigr)    \,,
 \nonumber \\
&&
{}^{(d+1)}R^{\mu}{}_{\alpha\nu\beta}
={}^{(d)}R^{\mu}{}_{\alpha\nu\beta}
-b'^2\Bigl(\delta^{\mu}{}_{\nu}\gamma_{\alpha\beta}
          -\delta^{\mu}{}_{\beta}\gamma_{\alpha\nu}\Bigr)  
 \nonumber \\
&&\hspace{1cm} +\frac{1}{2}
\Bigl(h_{\alpha}{}^{\mu}{}_{|\beta\nu}
    +h_{\beta}{}^{\mu}{}_{|\alpha\nu}
    -h_{\alpha\beta}{}^{|\mu}{}_{|\nu}
    -h_{\alpha}{}^{\mu}{}_{|\nu\beta}
    -h_{\nu}{}^{\mu}{}_{|\alpha\beta}
    +h_{\alpha\nu}{}^{|\mu}{}_{|\beta}
\Bigr)
 \nonumber \\
&&  \hspace{1cm}
-\frac{1}{2}b\,b'
  \Bigl(\delta^{\mu}{}_{\nu}h_{\alpha\beta}' 
       +\gamma_{\alpha\beta}h^{\mu}{}_{\nu}{}'
       -\delta^{\mu}{}_{\beta}h_{\alpha\nu}'
        -\gamma_{\alpha\nu}h^{\mu}{}_{\beta}{}'     
\Bigr)
 -b'^2\Bigl(\delta^{\mu}{}_{\nu}h_{\alpha\beta}
              -\delta^{\mu}{}_{\beta}h_{\alpha\nu}  \Bigr)
 \nonumber \\
&&\hspace{1cm}
-\frac{1}{2}
\Bigl[h^{\mu\rho}\Bigl(h_{\alpha\rho|\beta}
                      + h_{\beta\rho|\alpha}
                      -h_{\alpha\beta|\rho} \Bigr)\Bigr]_{|\nu}
+\frac{1}{2}
\Bigl[h^{\mu\rho}\Bigl(h_{\alpha\rho|\nu}
                      + h_{\nu\rho|\alpha}
                      -h_{\alpha\nu|\rho} \Bigr)\Bigr]_{|\beta}
 \nonumber \\
&& \hspace{1cm}
-\frac{b^2}{4}
\Bigl(h^{\mu}{}_{\nu}'h_{\alpha\beta}'
    -h^{\mu}{}_{\beta}'h_{\alpha\nu}' \Bigr)
 \nonumber \\
&&\hspace{1cm}
 -\frac{1}{2}b\,b'\Bigl(h^{\mu}{}_{\nu}{}' h_{\alpha\beta}
 - h^{\mu}{}_{\beta}{}' h_{\alpha\nu}
\Bigr)
+\frac{1}{2}b\,b'\Bigl(\gamma_{\alpha\beta}h^{\mu\rho}h_{\rho\nu}'
             -\gamma_{\alpha\nu}h^{\mu\rho}h_{\rho\beta}' 
   \Bigr)
 \nonumber  \\
&&\hspace{1cm}
+\frac{1}{4}
\Bigl(h^{\mu}{}_{\rho|\nu}h^{\rho}{}_{\alpha|\beta}
     +h^{\mu}{}_{\rho|\nu}h^{\rho}{}_{\beta|\alpha}
     -h^{\mu}{}_{\rho|\nu}h_{\alpha\beta}{}^{|\rho}
     +h^{\mu}{}_{\nu|\rho}h^{\rho}{}_{\alpha|\beta}
     +h^{\mu}{}_{\nu|\rho}h^{\rho}{}_{\beta|\alpha}
     -h^{\mu}{}_{\nu|\rho}h_{\alpha\beta}{}^{|\rho}
 \nonumber \\
&& \hspace{1cm}
     -h_{\rho\nu}{}^{|\mu}h^{\rho}{}_{\alpha|\beta}
     -h_{\rho\nu}{}^{|\mu}h^{\rho}{}_{\beta|\alpha}
     +h_{\rho\nu}{}^{|\mu}h_{\alpha\beta}{}^{|\rho}
     -h^{\mu}{}_{\rho|\beta}h^{\rho}{}_{\alpha|\nu}
     -h^{\mu}{}_{\rho|\beta}h^{\rho}{}_{\nu|\alpha}
     +h^{\mu}{}_{\rho|\beta}h_{\alpha\nu}{}^{|\rho}
 \nonumber \\
&& \hspace{1cm}
     -h^{\mu}{}_{\beta|\rho}h^{\rho}{}_{\alpha|\nu}
     -h^{\mu}{}_{\beta|\rho}h^{\rho}{}_{\nu|\alpha}
     +h^{\mu}{}_{\beta|\rho}h_{\alpha\nu}{}^{|\rho}
     +h_{\rho\beta}{}^{|\mu}h^{\rho}{}_{\alpha|\nu}
     +h_{\rho\beta}{}^{|\mu}h^{\rho}{}_{\nu|\alpha}
     -h_{\rho\beta}{}^{|\mu}h_{\alpha\nu}{}^{|\rho}
\Bigr).
\end{eqnarray}
The mixed components of the Ricci tensor are given by 
\begin{eqnarray}
&& {}^{(d+1)}R^{y}{}_{y}
          =-d\frac{b''}{b}
           +\frac{1}{4}h^{\rho\sigma}{}'h_{\rho\sigma}'
           +\frac{b'}{b}h^{\rho\sigma}h_{\rho\sigma}'  
           +\frac{1}{2}h^{\rho\sigma}h_{\rho\sigma}'' \,,
 \nonumber \\
&& {}^{(d+1)}R^{y}{}_{\nu}
          = -\frac{1}{2}h^{\rho\sigma}h_{\rho\nu|\sigma}'
            +\frac{1}{4}h^{\rho\sigma}{}_{|\nu}
                        h_{\rho\sigma}'
             +\frac{1}{2}h^{\rho\sigma}h_{\rho\sigma|\nu}'\,        
 \nonumber \\
&& {}^{(d+1)}R^{\nu}{}_{y}
          =\frac{1}{b^2}\Bigl(
             -\frac{1}{2}h^{\rho\sigma}h^{\nu}_{\rho|\sigma}{}'
            +\frac{1}{4}h^{\rho\sigma|\nu}
                        h_{\rho\sigma}'
             +\frac{1}{2}h^{\rho\sigma}h_{\rho\sigma}{}^{|\nu}{}'
              \Bigr)\,
  \nonumber  \\
&& {}^{(d+1)}R^{\alpha}{}_{\beta}
          = -\frac{b''}{b}\delta^{\alpha}{}_{\beta}
            +\frac{1}{b^2} {}^{(d)}R^{\alpha}{}_{\beta}
            -(d-1)\Bigl(\frac{b'}{b}\Bigr)^2\delta^{\alpha}{}_{\beta}
 \nonumber \\
&&  \hspace{1cm}
           -\frac{1}{2} h^{\alpha}{}_{\beta}{}''
           -\frac{1}{2}d \frac{b'}{b} h^{\alpha}{}_{\beta}{}' 
           +\frac{1}{2b^2}
        \Bigl(
                 h^{\alpha\rho}{}_{|\beta\rho}
                +h_{\beta}{}^{\rho|\alpha}{}_{|\rho}
                -\Box_{d}\, h^{\alpha}{}_{\beta} 
        \Bigr)    
         -\frac{1}{b^2}
            {}^{(d)}R_{\rho\beta}h^{\alpha\rho}
 \nonumber \\
&&\hspace{1cm}
            +\frac{1}{2} h^{\alpha\rho}{\,}'h_{\rho\beta}'
            +\frac{1}{2}\frac{b'}{b}\delta^{\alpha}{}_{\beta}
                h^{\rho\sigma}h_{\rho\sigma}'
            +\frac{1}{2} h^{\alpha\rho}h_{\rho\beta}''
            +\frac{1}{2}d\frac{b'}{b}h^{\alpha\rho}h_{\rho\beta}'
 \nonumber \\
&& \hspace{1cm}
            -\frac{1}{2b^2}h^{\rho\sigma}
           \Bigl(
              h_{\beta\rho}{}^{|\alpha}{}_{|\sigma}
             +h^{\alpha}{}_{\rho|\beta\sigma}
             -h^{\alpha}{}_{\beta|\rho\sigma}
             -h_{\rho\sigma}{}^{|\alpha}{}_{|\beta} 
           \Bigr)
           +\frac{1}{4b^2}
            \Bigl(
               h^{\rho\sigma|\alpha}h_{\rho\sigma|\beta}
             -2h^{\alpha\rho|\sigma}h_{\beta\sigma|\rho}
             +2h^{\alpha\rho|\sigma}h_{\beta\rho|\sigma}       
            \Bigr)
  \nonumber \\
&& \hspace{1cm}
          -\frac{1}{2b^2}h^{\alpha\rho}
           \Bigl(
              h_{\rho}{}^{\mu}{}_{|\beta\mu}
             +h_{\beta}{}^{\mu}{}_{|\rho\mu}
             -\Box_{d}\,h_{\rho\beta}
          \Bigr)
         +\frac{1}{b^2}{}^{(d)}R_{\gamma\beta}h^{\alpha\rho}
                               h_{\rho}{}^{\gamma}\,. 
\end{eqnarray}
The Ricci scalar is given by
\begin{eqnarray}
&& {}^{(d+1)} R= -2d\frac{b''}{b} 
      +\frac{1}{b^2}{}^{(d)}R 
      -d\bigl(d-1\bigr)\Bigl(\frac{b'}{b}\Bigr)^2
      -\frac{1}{b^2} {}^{(d)}R_{\rho\sigma}h^{\rho\sigma}  
 \nonumber \\
&& \hspace{1cm}
      +\frac{3}{4}h^{\rho\sigma}{}' h_{\rho\sigma}'
     +\bigl(d+1\bigr)\frac{b'}{b} h^{\rho\sigma}h_{\rho\sigma}'
     +h^{\rho\sigma}h_{\rho\sigma}''
     +\frac{1}{b^2}h^{\rho\sigma}\Box_{d}\,h_{\rho\sigma}
 \nonumber \\
&& \hspace{1cm}
  +  \frac{1}{b^2}
    \Bigl(
          \frac{3}{4}h^{\rho\sigma|\mu}h_{\rho\sigma|\mu}
         -\frac{1}{2}h^{\rho\sigma|\mu}h_{\rho\mu|\sigma} 
    \Bigr) 
    -\frac{1}{b^2}h^{\mu}{}_{\rho|\sigma\mu}h^{\rho\sigma}
    +\frac{1}{b^2}{}^{(d)}R_{\rho\sigma}
       h^{\rho}{}_{\mu}h^{\mu\sigma}\,.          
\end{eqnarray}
Using these results, the components of
the Einstein tensor are given by
\begin{eqnarray}
&& {}^{(d+1)}G^{y}{}_{y}
        =    -\frac{1}{2b^2}{}^{(d)}R
             +\frac{1}{2}d(d-1)\Bigl(\frac{b'}{b}\Bigr)^2
             +\frac{1}{2b^2} {}^{(d)}R_{\rho\sigma}h^{\rho\sigma}
  \label{bulkcurve}
\nonumber \\
&& \hspace{1cm} 
             -\frac{1}{8} h^{\rho\sigma}{\,}'h_{\rho\sigma}'
             -\frac{1}{2}(d-1)
                \frac{b'}{b}h^{\rho\sigma}h_{\rho\sigma}'
             -\frac{1}{2b^2} h^{\rho\sigma}\Box_{d}\,h_{\rho\sigma}
\nonumber \\
&& \hspace{1cm}
           -\frac{1}{2b^2}
            \Bigl(
            \frac{3}{4}h^{\rho\sigma|\mu}h_{\rho\sigma|\mu}
           -\frac{1}{2}h^{\rho\sigma|\mu}h_{\rho\mu|\sigma}
            \Bigr)
           +\frac{1}{2b^2}h^{\mu}{}_{\rho|\sigma\mu}
                       h^{\rho\sigma}
           -\frac{1}{2b^2}
              {}^{(d)}R_{\rho\sigma}h^{\rho}{}_{\mu}
                       h^{\mu\sigma}  
  \nonumber \\
&&  {}^{(d+1)}G^{y}{}_{\nu}={}^{(d+1)}R^{y}{}_{\nu} 
 \nonumber \\
&&  {}^{(d+1)}G^{\nu}{}_{y}={}^{(d+1)}R^{\nu}{}_{y}
 \nonumber \\
&&  {}^{(d+1)}G^{\alpha}{}_{\beta}
        = (d-1)\frac{b''}{b} \delta^{\alpha}{}_{\beta}
         +\frac{1}{b^2}{}^{(d)}G^{\alpha}{}_{\beta}
         +\frac{1}{2}(d-1)(d-2)\Bigl(\frac{b'}{b}\Bigr)^2
         \delta^{\alpha}{}_{\beta}
\nonumber \\
&& \hspace{1cm} 
         -\frac{1}{2}h^{\alpha}{}_{\beta}{}''
         -\frac{1}{2}d\frac{b'}{b}h^{\alpha}{}_{\beta}{}'
         +\frac{1}{2b^2}
        \Bigl(
           h^{\alpha\rho}{}_{|\beta\rho}
          + h_{\beta\rho}{}^{|\alpha\rho}
          -\Box_{d}\,h^{\alpha}{}_{\beta} 
        \Bigr)  
 \nonumber \\
&& \hspace{1cm}
       -\frac{1}{b^2}{}^{(d)}R_{\rho\beta}h^{\alpha\rho}
       +\frac{1}{2b^2}{}^{(d)}R_{\rho\sigma}h^{\rho\sigma}
           \delta^{\alpha}{}_{\beta}
\nonumber \\
&& \hspace{1cm}
     +\frac{1}{2} h^{\alpha\rho}{\,}'h_{\beta\rho}'
     +\frac{1}{2}h^{\alpha\rho}h_{\rho\beta}''
     +\frac{1}{2}d \frac{b'}{b}h^{\alpha\rho}h_{\rho\beta}'
     -\frac{1}{2b^2} h^{\rho\sigma}
         \Bigl(
            h_{\beta\rho}{}^{|\alpha}{}_{|\sigma}
           +h^{\alpha}{}_{\rho|\beta \sigma}
           -h^{\alpha}{}_{\beta|\rho\sigma} 
           -h_{\rho\sigma}{}^{|\alpha}{}_{|\beta}
         \Bigr)      
 \nonumber \\
&& \hspace{1cm}
     +\frac{1}{4b^2}
          \Bigl(
            h^{\rho\sigma|\alpha}h_{\rho\sigma|\beta}
            -2h^{\alpha\rho|\sigma}h_{\beta\sigma|\rho}
            +2h^{\alpha\rho|\sigma}h_{\beta\rho|\sigma}
          \Bigr)
    +\frac{1}{b^2}{}^{(d)}R_{\rho\beta}h^{\alpha\sigma}h_{\sigma}{}^{\rho}
 \nonumber  \\
&&  \hspace{1cm}
     -\frac{1}{2b^2}
         h^{\rho\alpha}
          \Bigl(
             h_{\rho}{}^{\sigma}{}_{|\beta\sigma}
            +h_{\beta}{}^{\sigma}{}_{|\rho\sigma}
            -\Box_{d}\,h_{\rho\beta}  
          \Bigr)  
 \nonumber \\
&& \hspace{1cm}
      -\frac{1}{2}\delta^{\alpha}{}_{\beta}
    \Bigl[
     \frac{3}{4}h^{\rho\sigma}{\,}'h_{\rho\sigma}'
    +d\frac{b'}{b}h^{\rho\sigma}h_{\rho\sigma}'
    +h^{\rho\sigma}h_{\rho\sigma}''
     +\frac{1}{b^2} h^{\rho\sigma}\Box_{d}\,h_{\rho\sigma}
\nonumber \\
&& \hspace{1cm}
    +\frac{1}{b^2}
      \Bigl(
          \frac{3}{4}h^{\rho\sigma|\mu}h_{\rho\sigma|\mu}
         -\frac{1}{2}h^{\rho\sigma|\mu}h_{\rho\mu|\sigma}
      \Bigr)  
    -\frac{1}{b^2}h^{\mu}{}_{\rho| \sigma\mu}h^{\rho\sigma}
    +\frac{1}{b^2}{}^{(d)}R_{\rho\sigma}h^{\rho}{}_{\mu}h^{\mu\sigma}
    \Bigr]\,. 
\end{eqnarray}

\section{Computational rules for averaging}

In this appendix, we describe the computational rules for averaging the
components of the second order part of the curvature tensors listed 
in Appendix~A.
 As we have noted in the main text, the notation $\langle A \rangle$
includes both the averaging along the ordinary spatial dimensions
 which are assumed
to be homogeneous and isotropic, and the small-scale 
time averaging as defined in
(\ref{timeavr}). In both cases,  
the computational rules are similar (See e.g., Ref. \cite{mtw}).
However, we do not apply the same rules for 
 terms with derivatives in the bulk direction,
because  we are dealing with a braneworld 
and the averaging along the bulk direction is ill-defined.

First, we note that we are interested in massive KK modes.
So, we can neglect terms coupled to the background curvature tensor as
\beq
            \Big  \langle
              {}^{(d)}R_{\rho\sigma} 
              h^{\rho}{}_{\mu}
              h^{\sigma}{}_{\nu}
            \Big  \rangle\,,\label{coup_BG}
\eeq
which are of order ${\cal O}(h^2/L^2)$, where $L$ is the
$d$-dimensional characteristic background curvature radius,
in comparison with terms as
\begin{eqnarray}
&&            \Big \langle
              h^{\rho\sigma}{}_{|\mu}
              h_{\rho\sigma|\nu}
            \Big \rangle\,,
       \quad
            \Big \langle
              h^{\rho\sigma}
              h_{\rho\sigma|\mu\nu} 
            \Big \rangle\,,
\nonumber \\
&&         \Big \langle
              h^{\rho}{}_{\mu}{}'
              h_{\rho\nu}'
           \Big \rangle \,,
       \quad
           \Big \langle
              h^{\rho}{}_{\mu}  
              h_{\rho\nu}''
           \Big \rangle \,,
         \cdots\,,
 \label{twice_diff}
\end{eqnarray}
which are of order ${\cal O}(m^2h^2)$. 
For instance for a cosmological brane with  expansion rate $H$,
we have $L={\cal O}(1/H)$. Thus $m\gg H$ implies $m\gg L^{-1}$, and we
can safely neglect corrections of the form (\ref{coup_BG}).

As a consequence, when taking the average, we are allowed to
freely interchange the order of the covariant derivatives.
For example,
\beq   \Big \langle 
          h^{\rho\sigma}{}_{|\mu\nu}
          h_{\rho\sigma}  
       \Big \rangle
      \simeq
        \Big \langle
          h^{\rho\sigma}{}_{|\nu\mu}
          h_{\rho\sigma}
        \Big  \rangle
          \,,  \label{commutative}
\eeq      
where corrections of order ${\cal O}(h^2/L^2)$ are neglected.
From now on, as in the main text we will use ``$=$'' instead of ``$\simeq$''
by neglecting the corrections.

Another computation rule is that total derivative terms can be neglected.
For example,
\beq
    \Big  \langle
        h^{\rho\sigma}{}_{|\mu}
        h_{\rho\sigma|\nu}
    \Big  \rangle
  = \Big  \langle
         \Bigl(
            h^{\rho\sigma}
            h_{\rho\sigma |\nu}
         \Bigr)_{|\mu}
     \Big \rangle    
   - \Big \langle
            h^{\rho\sigma}
            h_{\rho\sigma|\nu\mu} 
     \Big \rangle
 = - \Big \langle
            h^{\rho\sigma}
            h_{\rho\sigma|\nu\mu}
     \Big \rangle\,.  \label{move_diff} 
\eeq
This is because the total derivative term can be cast into the surface
integral which is smaller in magnitude
than the volume term by a factor $mR\,(\gg1)$,
where $R$ is the length scale of the averaging volume
 which is taken to satisfy $R\gg m^{-1}$.

As mentioned above, we do not apply 
the same rules for  terms with derivatives with
respect to the bulk coordinate $y$. However,   when one
considers  projections onto the brane, some simplifications occur.
On the brane, we have the boundary condition
$h_{\alpha\beta}'|_{\rm brane}=0$, which enables us to neglect
all the first derivative terms, e.g.,
\begin{eqnarray}
       \Big \langle
         h^{\alpha\rho}{}'h_{\rho\beta}{}'
       \Big \rangle\Bigr|_{\rm brane}
      = \frac{b'}{b}
       \Big \langle
         h^{\alpha\rho}h_{\rho\beta}{}'
       \Big \rangle\Bigr|_{\rm brane}
      =
       \Big \langle
         h^{\alpha\rho|\mu}h_{\rho\beta}{}'
       \Big \rangle\Bigr|_{\rm brane}
      =0\,.
\end{eqnarray}
In addition, using the bulk equation of motion~(\ref{eom}), 
we have
\begin{eqnarray}
   \Big \langle
        h^{\rho\sigma}\,h_{\rho\sigma}''
   \Big \rangle\Bigr|_{\rm brane}
  = -\Big \langle
        h^{\rho\sigma}\,\Box_{d}\,h_{\rho\sigma}
     \Big \rangle\Bigr|_{\rm brane}\,.
\end{eqnarray}

\end{document}